\documentclass[useAMS,usenatbib]{mn2e} %ASCII encoding required
\usepackage{amsmath}
\usepackage{graphicx}
\usepackage[T1]{fontenc}
\usepackage{epstopdf}
\title[]{Tidal-induced lopsidedness in Magellanic-type galaxies}
\author[C. Yozin and K. Bekki]{C. Yozin\thanks{{\bf E-mail} 21101348@student.uwa.edu.au; kenji.bekki@uwa.edu.au} and K. Bekki\footnotemark[1]
\\\\
ICRAR, M468, The University of Western Australia, 
35 Stirling Highway, Crawley
Western Australia, 6009, Australia}

\begin{document}
\date{Accepted 2013 January. Received 2013 January; in original form 2013 January}
\pagerange{\pageref{firstpage}--\pageref{lastpage}} \pubyear{2013}
\maketitle
\label{firstpage}

\newcommand{\ex}[1]{10$^{\rm #1}$}
\newcommand{\mex}[2]{#1$\times$10$^{\rm #2}$}
\newcommand{\rhoth}{$\rho_{\rm th}$}
\newcommand{\tsn}{$t_{\rm SN}$}
\newcommand{\rh}{r$_{\rm h}$}
\newcommand{\htw}{$H_{\rm 2}$}
\newcommand{\fgas}{f$_{\rm gas}$}
\newcommand{\md}{M$_{\odot}$}
\newcommand{\asym}[1]{A$_{\rm #1}$}
\newcommand{\ncm}{n$_{\rm H}$cm$^{\rm -3}$}
\newcommand{\hi}{H~{\sc I}}
\newcommand{\fkin}{f$_{\rm kin}$}
\newcommand{\kms}{kms$^{\rm -1}$}
\newcommand{\sfr}{M$_{\odot}$yr$_{\rm -1}$}
\newcommand{\esf}{e$_{\rm SF}$}
\newcommand{\sd}{M$_{\odot}$pc$^{-2}$}
\newcommand{\ha}{$H_{\alpha}$}
\newcommand{\bib}[1]{\bibitem[\protect\citeauthoryear{}{}]{} #1}

%%%% Figures
% 1 sep
% 2 profile
% 3 vcirc
% 4 disp
% 5 cont
% 6 asym1b
% 7 rho
% 8 asym1 
% 9 slope1b
% 10 asym5 
% 11 asym3

\begin{abstract}

We investigate the tidally-induced conversion of barred late-type spirals to Magellanic-type discs with numerical simulations, to establish how the lifetime of lopsidedness (asymmetry) varies with numerical parametrizations. Using a reference model based on observed properties of the Large Magellanic Cloud (LMC), we show that its off-centre bar, one-arm spiral and one-sided star formation can be induced by a brief tidal interaction. We thereafter perform a detailed parameter study, and find that at the dynamical mass of LMC-type discs ($\sim$10$^{\rm 10}$ \md{}), stellar lopsidedness (as quantified by the m$=$1 Fourier mode) and bar off-centredness can vary widely in amplitude, but are generally short-lived ($\sim$Gyr). Tidal interactions induce more persistent lopsidedness in lower mass galaxies (several Gyr), in particular those with large halo-to-disc mass ratios as implied by recent halo occupation models. We suggest that the tidal interactions play a larger role in the observed ubiquity of lopsidedness than the presently favoured gas accretion mechanism for lower mass galaxies. Other characteristics of Magellanic-type galaxies, such as one-sided star formation, are quantified and tend to arise more prominently in discs with later-type spiral structure (more halo dominated inner disc, weaker bars) following retrograde orbital encounters.

\end{abstract}


\section{Introduction}

The majority of spiral galaxies show substantial asymmetry (lopsidedness) in their stellar/gas morphology and kinematics (Richter \& Sancisi 1994; Matthews, van Driel \& Gallagher 1998). Those with an underlying barred late-type spiral structure (i.e. SBc-d) fall under the {\it Magellanic} classification of galaxies (de Vacouleurs \& Freeman 1972; Wilcots et al. 1996), as exemplified by the Large Magellanic Cloud (LMC). These include the one-armed NGC 4027 (Phookum et al. 1998) and NGC 4618, which possesses an off-centre bar with an extensive HI loop (Bush \& Wilcots 2004). The degree of lopsidedness is in many cases greater than that which can be explained by secular processes (Bournaud et al. 2005b; B05); the correlation in lopsidedness and star formation rate (SFR; Rudnick et al. 2000) would imply a mutual triggering event.

Tidal interactions/mergers have been long posited as the primary trigger of lopsidedness (Odewahn 1994; Zaritsky \& Rix 1997). This is consistent with recent surveys which suggest earlier-type spirals are generally more lopsided, partly attributable to such events in fact converting spirals to earlier types by means of bulge growth induced by tidal torquing (Elmegreen et al. 1990; Mihos \& Hernquist 1994). This argument may be valid in group environments (Angiras et al. 2006), where interactions occur frequently ($\sim$0.3 Gyr$^{\rm -1}$; Jog \& Combes 2009; and references therein). In contrast, many field/isolated disc with lopsided morphology have no clear evidence of companions/remnants (Wilcots \& Prescott 2004; B05) or are not bound to massive hosts (Odewahn 1994); the visible neighbour to Magellanic-type NGC 4618 for example may be too distant and kinematically distinct (Kaczmarek \& Wilcots 2012). 

Tidal triggers remain a viable progenitor if lopsidedness persists long after the interaction event (van Eymeren et al. 2011). Numerical simulations of merger/tidal-induced lopsidedness imply however that the corresponding lifetime lies on the order of a Gyr (Walker et al. 1996; B05), and is thus too short to account for its observed prevalence (Baldwin et al. 1980). B05 addresses this timescale problem by invoking gas accretion which, in idealised models, sustains lopsidedness for several Gyr. The galaxy sample which motivates their study finds later-type galaxies more lopsided however, in qualitative disagreement with more recent studies (i.e. van Eymeren et al. 2011). Moreover, their finding of a correlation in first and second Fourier modes (lopsidedness and bars/spirals respectively), an observation satisfied by accretion rather than a predominantly tidal trigger, conflicts with other samples of isolated galaxies (i.e. Espada et al. 2011).

To date, numerical studies have ascertained the lifetime of tidal-triggered lopsidedness in only limited detail. For example, disc structure is shaped by stellar processing: recent attempts to reproduce late-type spirals/bulge-less discs, such as the Galaxy, in simulations have invoked strong feedback (Governato et al. 2010), gas recycling from stars (Martig et al. 2010), long cooling times (Brook et al. 2004), or low efficiency of star formation (Agertz et al. 2011) to attain the requisite low bulge-to-disc ratios and halo profiles. The implications for lopsidedness have yet to be explored. 

Moreover, idealised models of lopsided galaxies raised the tantalising prospect of long-lived off-centre bars and global modes of asymmetry as a consequence of interactions/misalignments between the disc and halo (Jog 1997; Levine \& Sparke 1998; Noordermeer et al. 2001) and low pattern speeds (Ideta 2002; Saha et al. 2007). This is especially relevant to the broader issue of the dark matter content of disc galaxies; theoretical predictions based on halo occupation models (i.e. Moster et al. 2010) far surpass those obtained from rotation curves, for galaxies including the LMC (van der Marel \& Kallivayalil 2013).

The LMC presents an opportunity to benchmark simulated tidal-driven asymmetry against a wealth of observational data. Proper motions measurements and subsequent simulations (Besla et al. 2012; Diaz \& Bekki 2012) suggest the recent ($\sim$300 Myr ago) close passage by its less massive companion, the Small Magellanic Cloud, is primarily responsible for the LMC's exemplary lopsidedness. This has manifested in a single faint spiral arm and the offset of various subcomponents, including the optical centre of the bar and photometric centre of the outer isophotes, by $\sim$0.5 kpc (de Vacouleurs \& Freeman 1972; Westerlund 1997; Cole et al. 2005; van der Marel \& Kallivayalil 2013), which may be related to the elliptic and unvirialized nature of the disc (van der Marel 2001). The stellar bar is either warped (Subramaniam 2003), elevated above the disc plane (Zhao \& Evans 2000) or not a bar at all (Zaritsky 2004). The LMC also harbours its largest HII region (30 Doradus) at a single bar end, characteristic of Magellanic-type galaxies (Elmegreen \& Elmegreen 1980), and possibly related to the tidally-perturbed bar dynamics (Bekki \& Chiba 2007).

This paper investigates how tidal-induced lopsidedness in Magellanic-type galaxies depends on disc-halo interactions, galaxy mass, and parametrizations for star formation and feedback. The paper is organized as follows. In Section 2, we describe the parametrized numerical model. The main results for our reference models and a parameter study are given in Section 3, and the applicability of these results to a wider understanding of lopsided and Magellanic-type galaxies is discussed in Section 4.

\section{Numerical method}

\begin{figure}
\includegraphics[width=1.\columnwidth]{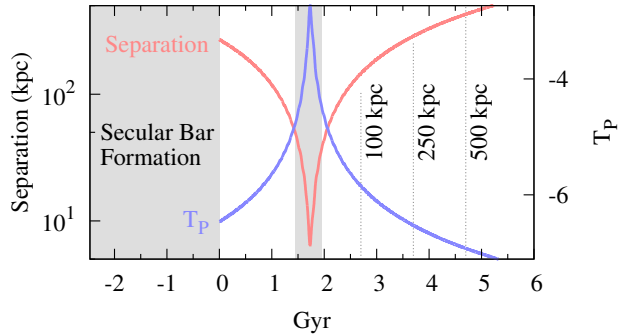}
\caption{Absolute distance between the primary and companion galaxies (red line) and the corresponding tidal parameter (T$_{\rm P}$; blue line). Our simulations commence with secular bar formation of the primary; the companion is added at T$=$0 Gyr on a hyperbolic orbit with pericentric distance of $\sim$7.5 kpc.} 
\label{fig_1}
\end{figure}

We simulate the response of a generic barred late-type spiral galaxy (the {\it primary}) to a brief tidal interaction with a less massive {\it companion}, using an original code (e.g. Bekki 2011). Accurate reproduction of differential tidal forces require self-consistent live models for both galaxies. The companion is structurally identical to the primary, except for its dynamical mass and mass-dependent metallicity. 

The primary's stellar bar is obtained first through an isolated preliminary simulation where a secular bar instability is induced in an isolated collisionless model (dissipative and SF physics are temporarily switched off) with a Toomre's parameter Q$=$1. This phase of the simulation is stopped when the disc has evolved to a quasi-stationary state (after several rotation times; Friedli \& Benz 1995) comparable to the LMC (namely the bar radius and disc scalelength; van der Marel 2001). Both dissipative physics, and the companion galaxy on a hyperboilc orbit of the primary, are then added and the simulation resumed at a reference time T$=$0 Gyr (Figure 1).

This method reduces the entanglement between the bar formation/growth phase (which is highly sensitive to parameters which subtly modify the disc self-gravity) and the subsequent interaction. One caveat lies in the bar density profile varying between tidally-induced and secular formation, with the latter being typically more cuspy (Noguchi 1996). The mass profile of bars in late-type galaxies indicate that they arise spontaneously, consistent with their relative isolation, and justifying our method. The typical epoch of bar formation for Magellanic-type discs is z$\sim$0.8-1 (Kraljic et al. 2012), thus prior to the recent epoch of LMC-SMC interaction. The apparent synchronicity of the LMC's bar formation with a close passage of the SMC (Smecker-Hane et al. 2002; Bekki \& Chiba 2005) however implies a tidal trigger. Alternatively, persistent tidal stripping of the halo by friction from the MW may have instead raised the relative self-gravity of the stellar disc leading to secular instability. Regardless, Figure 2 indicates that at T$=$0 the secular bar profile for our LMC-analogue is in fact relatively cored, with a clear cusp only developing later as consequence of tidally-induced infall. 

\begin{figure}
\includegraphics[width=1.\columnwidth]{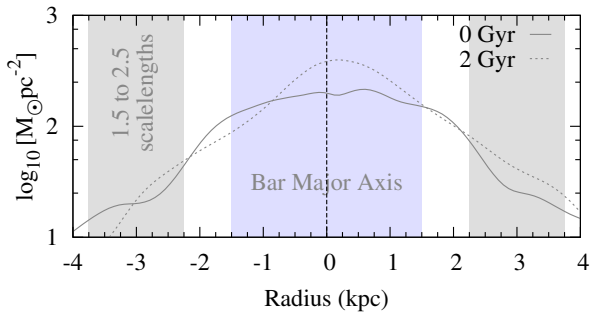}
\caption{Stellar surface mass density along the bar major axis at commencement of the reference retrograde simulation (0 Gyr) and several Myr following pericentre (2 Gyr). The bar radius (1.5 kpc) and annulus in which global asymmetry is determined are shaded blue and grey respectively.} 
\end{figure}

\subsection{Interaction Parameters}

We model our tidal interaction on the most recent pericentre in the binary orbits of the LMC/SMC, occurring 300 Myr ago (Kallivayalil et al. 2006; Diaz \& Bekki 2012); the previous close interaction with the SMC (and MW) lies at least 1.5 Gyr earlier. The pericentric distance r$_{\rm p}$ is accordingly set to 8 kpc. We take the null hypothesis that the LMC's myriad asymmetry is a consequence of this recent interaction; B05 show that {\it strong} asymmetry correlates with the perturbation of the primary's potential in a single Gyr period following pericentre. 

We use a primary-to-companion dynamical mass ratio \rh{} of 0.3, consistent with the models of Diaz \& Bekki (2012) which successfully reproduce the tidal formation of the Magellanic Stream over the previous 2 Gyr. The strength of the interaction can be quantified with a tidal parameter (T$_{\rm P}$):
\begin{equation}
T_P = log\left( r_m\left( \frac{R_S}{r_p} \right)^3 \right),
\end{equation}
where r$_{\rm s}$ is the stellar disc scalelength (Section 2.3). Our prescribed tidal scenario corresponds to a peak T$_{\rm P}$ of about -2.8 and a relative velocity of $\sim$250 \kms{}; the corresponding peak lopsidedness (\asym{I}{}; Section 2.4) of $\sim$0.2 is thus consistent with Fig. 11 of B05 which considers asymmetry over a range of T$_{\rm P}$.

The absence in correlation between lopsided discs and visible companions (Wilcots \& Prescott 2004; B05) motivates us to implement a highly eccentric hyperbolic orbit for the companion. Magellanic irregulars are frequently isolated although a quantification of {\it isolation} in this case is generally lacking. For the specific case of asymmetric Magellanic galaxies NGC 4625 and 4618, which are of comparable mass to the MCs but separated by $\sim$900 kpc; Bush \& Wilcots (2004) estimate, on the basis of an HI loop in NGC 4628, a most recent passage on the order of 1 Gyr ago. Under a simple two-body orbital scenario, this requires an exceptionally large eccentricity ($e>$50). 

We take a moderate approach more consistent with the MCs, wherein $e=$5 provides a 500 Myr window during which the companion lies within the primary's tidal radius of $\sim$20 kpc (van der Marel 2002), and is 0.1 and 0.5 Mpc distant at 1 and 3 Gyr after pericentre respectively. Other prescriptions for isolation among massive spirals similarly require no neighbour within several hundred kpc (Tollerud et al. 2011; Lorenzo et al. 2013), or 20 times the primary's angular diameter (Verley et al. 2007). Moreover, B05 suggest that a tidal parameter of T$_{\rm P}$$<$-6 represents a very isolated state. No significant tidal structures bridging our model galaxies form; based on the aforementioned criteria therefore, the primary galaxy could be feasibly identified as isolated only 1 to 1.5 Gyr after pericentre.

We mainly consider models wherein the companion orbits the primary in retrograde (the orbit trajectory is counter to the primary disc rotation), and enforce an angular offset (phase-lag) between the primary bar and companion of approximately 30 to 60 degrees to reduce degeneracy introduced by potentially polluted resonant responses (Gerin et al. 1990). The companion disc is coplanar with its orbital plane, which itself is inclined to the primary disc by some inclination angle $i$. Our reference model uses i$=$45 degrees, to introduce the tidal resonant pollution expected across interacting galaxies for which we assume no general bias towards the innately stronger zero inclination case

\subsection{Disc Model}

\begin{figure}
\includegraphics[width=1.\columnwidth]{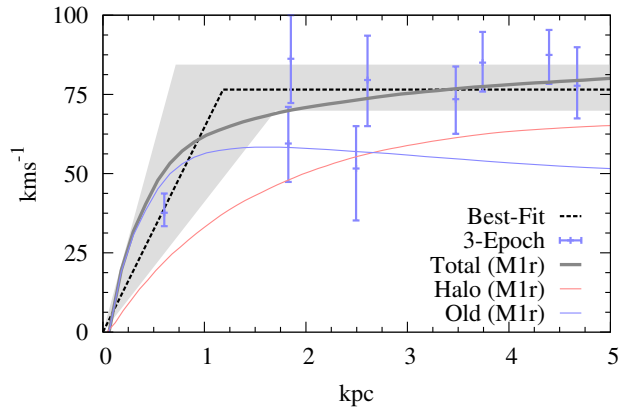}
\caption{Rotation curve of the primary reference model following the orbital pericentre, assuming v$=$$\sqrt{GM(<r)/r}$, for old stars, the halo and all components (blue, red and grey lines respectively). Recent 3-epoch observational data from Van der Marel \& Kallivayalil (2013) is conveyed with blue points with their line-of-best fit (black line) with 1$\sigma$ uncertainty (grey region).} 
\end{figure}

The bulge-less stellar and gas discs have an initial total mass M$_{\rm d}$, and lie embedded within a massive halo with mass $M$. The halo mass is defined in terms of the disc mass with ratio r$_{\rm h}$. In the case of the LMC and analogues, predictions for the halo mass fall into two groups. Direct observations of stellar/HI kinematics  determine r$_{\rm h}$ on the order of 1 to 10 (Bush \& Wilcots 2004; Diaz \& Bekki 2012, van der Marel \& Kallivalyalil 2013). Statistical methods applied to cosmological models, however, appear to raise the typical ratio by up to a magnitude. Moster et al. (2010) for example equate the of-order \ex{9}{} \md{} baryonic mass of the LMC with a halo mass \mex{0.5}{11}{} \md{}. Boylan-Kolchin, Besla \& Hernquist (2011) consider mass {\it and} environment for LMC-analogues and determine a similar estimate. 

We adopt M$=$\mex{1.7}{10}{} \md for our reference model, in accord with van der Marel \& Kallivayalil (2013), corresponding to r$_{\rm h}$$\sim$4, and larger than previous detailed models of the LMC, such as Bekki \& Chiba (2005) who adopt a halo-to-disc mass ratio for their models of 2.33 consistent with van der Marel et al. (2002). We utilise the fixed halo density profile suggested by Salucci \& Burkert (2000) for dwarf galaxies:
\begin{equation}
{\rho}_{\rm dm}(r)=\frac{a_{\rm h}\rho_{\rm dm,0}}{(r+a_{\rm dm})(r^2+{a_{\rm dm}}^2)},
\end{equation}
where $\rho_{\rm dm,0}$ is the central dark matter density, and a$_{\rm h}$ is a free scaling parameter. The fundamental parameters for this dark matter profile $\rho_{dm,0}$, $a_{\rm dm}$, and M$_{0}$ (where M$_{0}$ is the total dark matter mass within $a_{\rm dm}$) are not free parameters, and clear correlations are observed between them (Burkert 1995):
\begin{equation}
M_{0}=4.3 \times 10^7 {\left(\frac{a_{\rm dm}}{\rm kpc}\right)}^{7/3} M_{\odot}.
\end{equation}

The observed LMC stellar mass (\mex{2.7}{9}{} \md; van der Marel et al. 2002) neutral gas mass (\mex{5}{8}{} \md; Kim et al. 1998) and presumed gas loss by SF and tidal-stripping motivate a choice of M$_{\rm d}$$=$\mex{3.5}{9}{} \md, with a corresponding gas mass fraction \fgas{} of 30 percent. The truncated stellar disc size R$_{\rm d}$ is assumed to scale with a size-luminosity power law, normalised by Galactic values: 
\begin{equation}
\frac{r_d}{17.5 kpc}=\left(\frac{M_d}{6\times10^{10}M_\odot}\right)^{0.32},
\end{equation}
The radial ($R$) and vertical ($Z$) density profiles of the stellar disc are assumed to be proportional to $exp(-R/R_{0})$ (with scale length $R_{0}$$=$0.2R$_{\rm d}$) and ${\rm sech}^2 (Z/Z_{0})$ (with exponential scale length $Z_{0}$$=$0.04R$_{\rm d}$) respectively. This provides a R$_{\rm 0}$ consistent with the LMC's observed 1.3 kpc (van der Marel et al. 2002). The exponential gas component is modelled with a single phase, and extends to at least 1.5 times the stellar disc radius. In addition to the rotational velocity caused by the gravitational field of disc and dark halo components, the initial radial and azimuthal velocity dispersions are assigned to the disc component according to the epicyclic theory with Toomre's parameter $Q$$=$1.

For the aforementioned parameter set, a$_{\rm h}$ was tuned to find a model rotation curve (assuming V$=\sqrt{GM(<r)/r}$) that complies with observed data shortly after pericentre (Figure 3). If interpreted in terms of a two component profile, the data implies a steep rise to approx 76 \kms{} at 1.4 kpc, and remains largely constant at greater radii (Luks \& Rohlfs 1992; van der Marel \& Kallivayalil 2013). This profile is consistent with late-type spirals, with the bar radius (1.5 kpc in the case of LMC) often constrained by the turnover radius (Combes \& Elmegreen 1993). Van der Marel \& Kallivayalil (2013) note that the LMC lies comfortably within the tight Baryonic Tully-Fisher relation, which the implication that tides and the bar are not substantially modifying the rotation curve over time.

\begin{figure}
\includegraphics[width=1.\columnwidth]{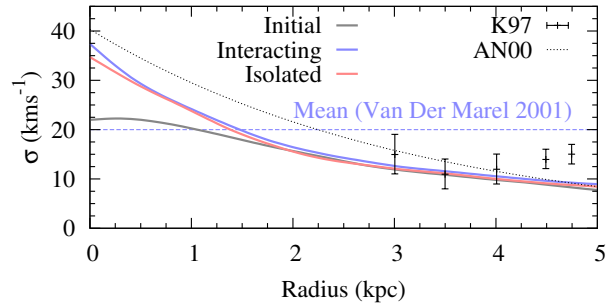}
\caption{Vertical velocity dispersion as a function of radius at T$=$0 Gyr (grey line), and T$=$2 Gyr for the isolated and retrograde simulations (re and blue lines respectively). The analytical profile of Alves \& Nelson (2000; AN00), derived from estimates of the present disc structure, is conveyed with the grey dashed line; the mean observed $\sigma$ of van der Marel (2001) and data points from Kunkel et al. (1997; K97) are given with blue dashed line and black error-bars respectively.}
\label{fig_6}
\end{figure}

Figure 4 shows the vertical velocity dispersion as a function of radius, in isolated and interacting (retrograde) orbital models at T$=$2 Gyr (shortly after pericentre in the interaction case). Disc stars are heated perpendicular by the bar instability, leading to a thick discs similar to those observed among Magellanic-type galaxies (de Vacouleurs \& Freeman 1972); the heating due to the weak inclined tidal interaction (Quinn et al. 1993) is only minor by comparison. Both models are consistent with carbon star kinematics (Kunkel, Irwin \& Demers 1997), measurements for other subpopulations which spread from 6 to 30 kms$^{\rm -1}$ (Van der Marel 2001 and references therein); and an analytical dispersion profile based on present estimates of R$_{\rm 0}$ and Z$_{\rm 0}$ (Alves \& Nelson 2000).

Our choice of structural parameters generally lead to the formation of an Inner Lindblad Resonance (ILR), indicated by gas lying in a nuclear concentration and circum-nuclear ring (i.e. Knapen et al. 1995), consistent with spiral discs in general. As a specific case, the LMC is notably more homogeneous in its  \hi{} distribution (Staveley-Smith et al. 2003) and lacking a gaseous bar, unlike other Magellanic-type disc (Wilcots et al. 1996). We presume the unique nature of the long-term interaction with the SMC and/or truncation by the MW halo (Mastropietro et al. 2005), which we do not model here, accounts for these differences.

\begin{figure}
\includegraphics[width=1.\columnwidth]{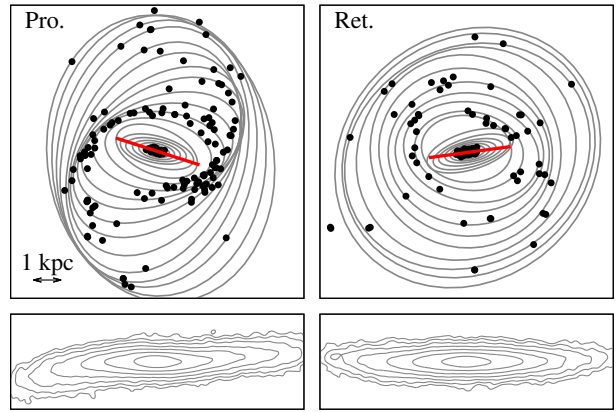}
\caption{(Top) Iso-density ellipses (grey lines), the stellar bar (red line) and young stars (age less than 300 Myr; black points). The left and right panels show the prograde and retrograde reference models respectively, 500 Myr after pericentre, by which time the characteristic morphologies of these interactions have developed; (Bottom) A 1 kpc-wide cross-section viewed edge-on with the x-axis representing the bar major axis. Logarithmic scale contours convey stellar surface density.} 
\end{figure}

\subsection{Star Formation and feedback}

A gas particle is converted into a new star with efficiency \esf{} if (i) the local dynamical time-scale is shorter than the sound crossing time scale (mimicking the Jeans instability), (ii) the local velocity field is identified as being consistent with gravitationally collapsing (i.e. div {\bf v}$< 0$), and (iii) the local volumetric density exceeds a threshold density for star formation (\rhoth{}). We fine-tune the SF criteria to match the observed star formation rate of order 0.1 \sfr{} typical of the LMC's recent past (Harris \& Zaritksy 2009), and to account for mass resolution dependence. Figure 6 exhibits how the tidal interaction can be clearly imprinted on the star formation history, only $\sim$1 Gyr after a burst corresponding to the initial collapse of the gas component. This motivates our choice for a reference \rhoth{} as 10 \ncm{}, with a star forming efficiency of 1.5 percent consistent with observation (i.e. Krumholz \& Tan 2007). We adopt a Salpeter IMF with new stars in the mass range 0.1 \md{} $<$ M$ <$ 50 \md{}.

\begin{figure}
\includegraphics[width=1.\columnwidth]{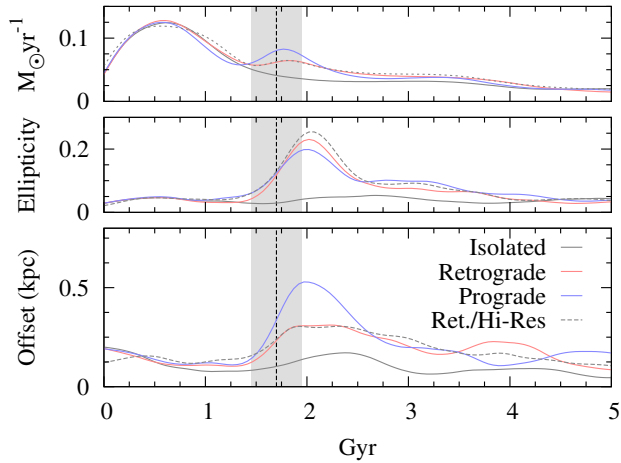}
\caption{(Top) Star formation history of the reference isolated, prograde and retrograde (reference- and high-resolution models); (Middle) Ellipticity of the stellar disc. Van der Marel (2001) finds the LMC possesses a value of 0.2 to 0.3; (Bottom) In-plane separation (offset) between the disc and bar centres; the observed LMC bar exhibits an offset of 0.4 to 0.5 kpc (van der Marel 2001).}
\end{figure}

Supernova feedback is implemented at the sub-grid level on the assumption that all interacting energetic output (a canonical \ex{51}{} ergs) from the event is shared between thermal (UV emission) and mechanical kinematic components; the balance is controlled with the \fkin{} parameter, with 0 and 1 corresponding to pure thermal. The shock extends to a radius of 0.175 kpc, and thermal transfer is enforced over a period of 100 Myr. In a cursory attempt to match the underlying morphology, SFR and age-metallicity relation for the quiescent epoch in the LMC's history (e.g. from Cole et al. 2005, Harris \& Zaritsky 2009), there arises a need to eject the majority of SN energy into thermal form (i.e. \fkin{}$=$0.1). This prescription also prevents spurious clumps in the gas distribution for a gas mass fraction in the range 0.2 to 0.3 as applicable here, and maintains an extended disc. This range of \fkin{} is consistent with observations of SNe (Korpi et al. 1998), and earlier simulations (Navarro \& White 1993). Radiative cooling, with $\eta = 5/3$, follows the rate curves of Rosen et al. (1993) in the range $10^2 < T(K) < 10^4$, and Mappings III (Allen, Sutherland \& Dopita 1993) for $T > 10^4 K$. 

\subsection{Quantifying lopsidedness}

Fitting equal-mass ellipses (Pilu, Fitzgibbon \& Fisher 1999) to a smoothed face-on surface mass density distribution provides a robust means of calculating the disc centre, which is taken as an the average of the centres of the outermost ellipses. The central-most ellipse with eccentricity$>$0.5 is taken as the bar centre and, with respect to this location, the radial-azimuthal mass distribution of the disc is obtained. Using the method of Odewahn et al. (2002), a Fourier decomposition determines the zeroth to second mode amplitudes/phase. 

We use the empirical measure of disc-wide asymmetry \asym{I}{} suggested by Zaritsky \& Rix (1997). The scalelength r$_{\rm s}$ is estimated to coincide with one $e$-folding less than the central luminosity attained from extrapolating the radial profile from the half-luminosity radius. \asym{I}{} is then the average M$=$1 amplitude, normalised by the corresponding M$=$0 amplitude, in the range 1.5 to 2.5r$_{\rm s}$ from the luminous centre (Figure 2). Note that we assume for brevity a constant mass-luminosity relationship, and that our face-on \asym{I}{} can be compared with observations in spite of inclination uncertainty.

The bar radius is defined as the shorter of (i) the radius where the m=2 phase starts to deviate from its established phase by more than 10 degrees (e.g. O`Neill \& Dubinski 2003) and (ii) the radius where the corresponding ellipse minor-major axis ratio starts to exceed 0.4. We find in practice these to be robust criteria (Figure 5). The in-plane separation between the disc and bar centres is used as a measure of bar offset within the host halo; Figure 6 conveys how our reference prograde orbital model attains an offset (and disc ellipticity) consistent with the observed $\sim$0.5 to 0.6 kpc (0.2 to 0.3) of the LMC (van der Marel 2001). Bar strength is quantified as the maximum amplitude of the m$=$2 mode within the bar (Athanassoula et al. 2013).

\subsection{Numerical Resolution}

The following analysis is predicated on the plausible response of the primary halo to external perturbations, and its relationship with the inner disc and adiabatic compression by the stellar bar. This mechanism, like others concerning the inner halo, depends strongly on model resolution (e.g. Theis 1997; Weinberg 1998; Athanassoula \& Msirotis 2002). Our adoption of 500,000 particles for the primary galaxy (with the allocation of 400,000 to the halo, 100,000 to the disc) lies within the recommended ranges of the aforementioned studies. The number of particles describing the companion galaxy depends on the mass ratio; for a reference r$_{\rm m}$$=$0.3, our simulations thus have a total of 650,000 particles. 

Moreover, our models use a variable time-step ranging from \mex{1.4}{4}{} to \mex{1.4}{6}{} on the order of 40 kyr (depending on particle density), which is consistent with the convergent regime range identified by Dubinski, Berentzen \& Shlosman (2009) to yield convergent bar and halo central densities. Other studies indicate however that \ex{6}{} particles cannot resolve the interaction between the bar and halo (Holley-Bockelmann et al. 2005). 

\begin{figure}
\includegraphics[width=1.\columnwidth]{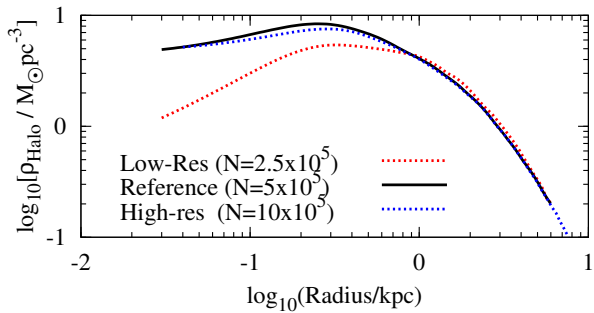}
\caption{Volumetric halo density as a function of radius from the dynamical centre, for various particle numbers $n$ of the primary galaxy. The reference- and high- resolution models are sufficiently converged within the inner kpc.} 
\end{figure}

The incorporation of a dissipative medium and sub-grid physics in a parameter study format limits the resolution we can adopt in practice. Consequently, we compare the salient results of our retrograde reference model with those of a high resolution case (1,300,000 particles in total for r$_{\rm m}$$=$0.3). Figures 5 and 8 indicate that disc-wide metrics of star formation and lopsidedness are acceptably consistent in both reference- and high-resolution cases for the retrograde orbital scenario. Moreover, halos tend to flatten due to torquing by the bar (Holley-Bockelmann et al. 2005), but excessive flattening is characteristic of poor spatial resolution (Dubinski et al. 2009), as exemplified in our low-resolution case (Figure 7); the reference case is however acceptably converged to the higher-resolution results.

\section{Results}

Table 1 summarises the structural and physical parameters for reference simulations we find consistent with the LMC's observable benchmarks. The results for a one high resolution and 26 reference resolution models are discussed in the following section. 

\begin{table} %*}
\centering
%\begin{minipage}{100mm}
\caption{Summary of reference model (M1) parameters}
\begin{tabular}{@{}ccc@{}}
\hline
Parameter & Value & Definition \\
\hline
$N$ & 500,000 & Particle Number \\
$M$ & 17$\times$10$^{\rm 9}$ M$_{\odot}$ & Dynamical Mass  \\
$M_{\rm d}$ & 3.5$\times$10$^{\rm 9}$ M$_{\odot}$ & Baryonic Mass \\
f$_{\rm gas}$ & 0.3 & Initial Gas Fraction \\
$r_{\rm s}$ & 1.4 kpc & Initial stellar scalelength \\
$z_{\rm s}$ & 0.28 kpc & Initial stellar scaleheight \\
$r_{\rm g}$ & 4.2 kpc & Initial gas scalelength \\
$z_{\rm g}$ & 0.56 kpc & Initial gas scaleheight \\
Q & 1 & Initial Toomre parameter \\
a$_{\rm h}$ & 0.75 & Halo density scaling \\
f$_{\rm kin}$ & 0.1 & Kinematic feedback factor \\
t$_{\rm SN}$ & \mex{4}{4}{} yr & SNe expansion timescale \\
$\rho_{th}$ & 10 ncm$^{\rm -3}$ & SF Threshold Density \\
$e$ & 0.015 & SF efficiency \\
$r_{\rm m}$ & 0.3 & Mass ratio \\
$r_{\rm p}$ & 8 kpc & Pericentre of two-body orbit \\
$\theta$ & 45 degrees & Orbit inclination \\
$e$ & 5	 & Orbit eccentricity \\
\hline
\end{tabular}
%\end{minipage}
\end{table} %*}

\subsection{Reference Model}

\subsubsection{In-plane morphology}

Figure 8 compares the time evolution of \asym{I}{} in reference models with prograde/retrograde companion orbits, and an isolated (no companion galaxy) simulation. The interaction scenarios clearly enhance lopsidedness; the isolated case is consistent with B05 who find intrinsic mechanisms can not promote \asym{I}$>$0.05. The prograde case results in {\it strong} lopsidedness, which is defined by the top 20 percent of a sample of galaxies considered by Rudnick \& Rix (1998), which exhibit \asym{I}$>$0.19. By contrast, they find a median of 0.11, consistent with more recent samples (0.1; van Eymeren et al. 2011). In both interaction cases, the timescale for above-average \asym{I}{} lies on the order of a Gyr, or about 10 dynamical times. This is consistent with the rapid decline in T$_{\rm P}$ (Figure 1; and B05), and rapid dissipation of bulk stellar asymmetry through differential disc rotation or damping by a density wake in the halo. We find in these and later models that the bar offset (in-plane separation between bar and disc centres) correlates with \asym{I}; the prograde and retrograde cases exhibit peak offsets of 0.5 and 0.3 kpc respectively, broadly consistent with the LMC (0.4 to 0.5 kpc; van der Marel 2001).

Based on the present rotation of the LMC (van der Marel et al. 2013) and recent simulations constrained by proper motions (Diaz \& Bekki 2012), the most recent LMC-SMC close passage conforms to a retrograde orbital encounter. Figure 5 indicates that a prograde encounter by contrast enhances the spiral arms, traced in particular by recent star formation along the associated arms; the retrograde case shows no such large scale SF distribution, consistent with the LMC. The prograde encounter induces a stronger \asym{I}, because the companion's relative velocity ($\sim$250 \kms{}) at pericentre, being far greater than the primary's rotation velocity, preferentially perturbs/strips one side of the disc. This is qualitatively different from Bournaud et al. (2005a) who find, in models with a lesser relative velocity compared to rotation, that the retrograde case yields stronger \asym{I}{} in mergers because the companion is less deformed prior to pericentre and thus exerts a more coherent tidal force.

In both pro/retrograde cases, the {\it star-burst} attributable to tidal-induced infall achieves little more than a doubling of the corresponding secular rate, consistent with the disc-wide SFH of the LMC (Harris \& Zaritsky 2009), and simulations of gas-rich interacting galaxies (Di Matteo et al. 2007). The last 5-6 Gyr show an underlying quiescent rate punctuated by brief doubling episodes which appear to coincide with the close passages of the SMC. For a brief period of several 100 Myr (consistent with Di Matteo et al. 2007), the prograde case exhibits a more enhanced SFR because gas overdensities in the spiral arms and nucleus are more optimal for SF than the smaller stochastic overdensities formed in the retrograde case. The retrograde model retains only a weak spiral structure, again preferentially enhanced on one side, and which does not harbour any strong SF (Figure 5). By a Gyr after pericentre, when lopsidedness in both cases has largely subsided, the disc morphologies as traced by the exponential disc scalelength and bar size are similar, suggesting that this high speed encounter does not promote transformation to earlier type spirals (Elmegreen et al. 1990).

The disc ellipticity in both interacting reference models is similar (Figure 6), rising to the 0.2-0.3 traced by RR Lyrae in the LMC (van der Marel 2001), which are generally smoothly distributed (Hashke et al. 2012) and trace the underlying triaxiality of the distorted halo. This large scale elongation of the stellar disc, exceeding the average among spirals, has in previous studies been associated with the long-term influence of the Galaxy (Weinberg 2000; van der Marel 2001). 

\begin{figure}
\includegraphics[width=1.\columnwidth]{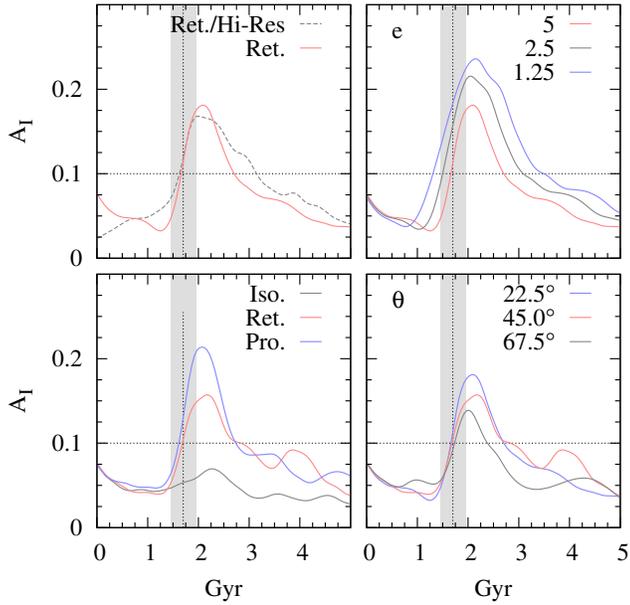}
\caption{Asymmetry of the isolated, prograde and retrograde- (and high-resolution retrograde) reference models; the shaded region indicates the epoch during which the companion lies within the tidal radius of the primary. (Top left) Reference and High-resolution retrograde models; (Bottom left) Isolated, prograde and retrograde reference models; (Top right) reference retrograde model with varying orbital eccentricity of companion (reference $e$$=$5); (Bottom right) reference retrograde model with varying orbital inclination (reference inclination 45$^{\circ}$).} 
\end{figure}

\subsubsection{SF and out-of-plane lopsidedness}
% off sf

The distorted outer disc and brief tidal torquing leads to uneven gas infall to the bar. The noted presence of an ILR prior to the interaction (Section 2.2) does not mask this effect in the inner disc, when quantified in terms of one-sided star formation (due to offset shocks on the leading edge of bar caused by the collision of infalling gas and gas bound to the resonant stellar orbits of the bar). Figure 9 shows the mass-weighted mean radius of recent star formation (less than 100 Myr) parallel to and normalised by the the bar length (r$_{\rm SF}$). The data is implicitly averaged in time (with data output from the simulations at 70 Myr increments, compared to typical GMC's lifetimes of $\sim$20 Myr i.e. Murray 2010).

The retrograde encounter induces a larger r$_{\rm SF}$, in a 2 Gyr period following pericentre, than in the isolated case. This can be attributed to the underlying lopsided potential; Jog (1997) showed how overdense gas at the disc azimuth of peak M$=$1 can exhibit enhanced SF of up to 50 percent. The significantly smaller r$_{\rm SF}$ in the prograde case is indicative of a qualitatively different response in the inner disc, in spite of a greater overall \asym{I}: the encounter promotes stronger infall via an enhanced m$=$2 mode, with the subsequent growth of a centrally concentrated bulge promoted over stochastic SF at the bar ends.

The perturbed potential can manifest in out-of-plane structure, which we propose here to be related to one-sided SF. These structures, arising only in non-isolated models, include flaring of the disc, tidal arms tending towards the inclined path of the perturber's orbit and misalignment of the bar. The bar of our retrograde model is both mis-shapen and shows the strongest misalignment relative to the disc plane among our reference models (Figure 9). The timescale over which it surpasses that of the isolated model is also consistent with r$_{\rm SF}$. 

Dereddened magnitudes taken along the LMC bar indicate that it is warped, with the bar ends of the nearly face-on disc lying closest from our perspective (Subramaniam 2003). The strong deformation may also explain observations of the bar, which suggest a substantial displacement from the disc plane or even a misaligned bulge (Zhao \& Evans 2000; Zaritsky 2004). Besla et al. (2012) find in tidally-perturbed models of the LMC a significant bar warp (the inclination of the bar relative to the disc), which they suggest is indicative of a weak bar and explains the near absence of a gaseous counterpart to the stellar bar (Kim et al 1998). 

Out-of-plane distortions in the prograde model manifest more prominently as warping of the outer disc (Figure 5), which we quantify with the maximum mid-plane displacement from the disc plane (Figure 9). This warp is short-lived, comparable to the 1 Gyr timescale of T$_{\rm P}$ and \asym{I}. Such warps are common among spirals, preferentially occurring beyond several scalelengths as we find here (Saha \& Jog 2006), but are not necessarily related to lopsidedness (Jog \& Combes 2009). 

\begin{figure}
\includegraphics[width=1.\columnwidth]{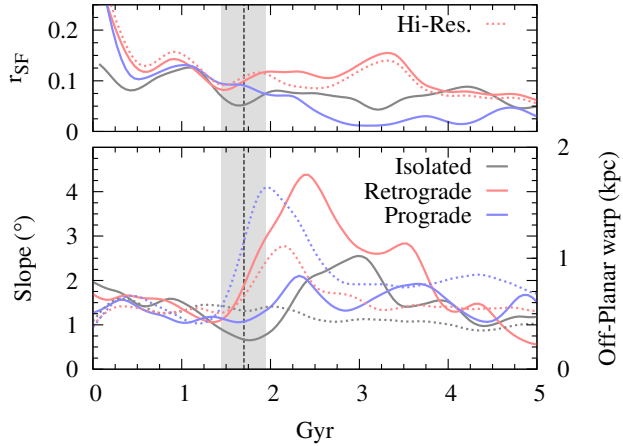}
\caption{(Bottom) For the same models, bar warp angle (thick lines) and disc warp (dashed lines); (Top) Mean radii of star formation, normalised by the bar radius, as a function of simulation time, for the isolated, prograde and reference- and high-resolution retrograde models}
\end{figure}

\subsection{Dependence on orbital parameters}

The reference retrograde model of Section 3.1 most closely reproduces the salient aspects of the benchmark LMC, so we hereafter perform a parameter study based on this interaction scenario. For our reference T$_{\rm P}$ at pericentre, inclination and eccentricity are the orbital variables most influential on lopsidedness (Bekki 2009). Reducing the eccentricity allows for a larger T$_{\rm P}$ over a longer period of time, with the tidal force itself enhanced at the lower corresponding transit velocity (Figure 9). The lower eccentricities considered are not compatible with the primary appearing isolated within a short period after the interaction, unless the companion is a dark halo (Bekki 2009), diffuse gas cloud (Wilcots et al. 1996), or is otherwise too low mass for detection, in which case a lower $e$ could compensate for the reduced mass ratio/T$_{\rm P}$ (Section 3.5). In this latter case, the deformation of the companion is significant, such that retrograde encounters (in which the companion is deformed less) yield greater \asym{I}, in accord with Bournard et al. (2005a) but discrepant from our reference scenario.

We find the most co-planarity between orbit and primary disc (22.5 degrees) yields the largest \asym{I}{} (Figure 8), due to a largest in-plane component to the tidal force. On the other hand, the bar warp and r$_{\rm SF}$ are negligible; instead, the stellar bar establishes a strong boxy/peanut-shape, characteristic of an induced vertical buckling, that substantially weakens the bar (Martinez-Valpuesta \& Shlosman 2004) but remains in-plane with the disc. By contrast, the 45 and 67.5 degree inclination models establish weak bars with either a systematic warp or bow-shape through their length; r$_{\rm SF}$ is accordingly larger (equally so in both models due a poorer dependence towards larger inclinations). The observed LMC bar is more consistent with the latter cases (Subramaniam 2003).

\subsection{Dependence on galaxy mass}

\begin{figure}
\includegraphics[width=1.\columnwidth]{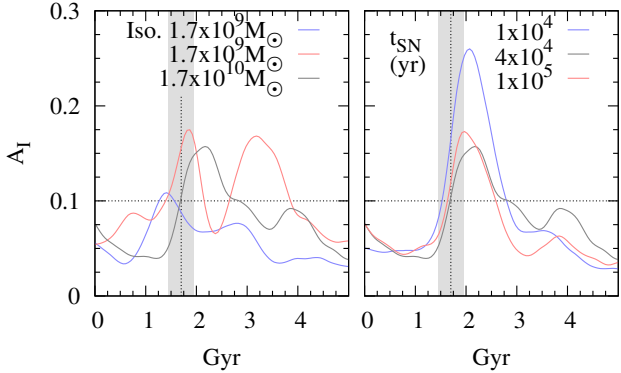}
\caption{Same as Figure 8 for (Left) Reference retrograde model, low-mass retrograde and isolated model; (Right) varying adiabatic timescale t$_{\rm SN}$ for SNe (reference t$_{\rm SN}$$=$4x10$^{\rm 4}$ yrs).} 
\end{figure}

The relative paucity of low mass halos relative to theoretical expectation (Moore et al. 1999) is thought to be due to a weak potential well in which stellar feedback and external perturbations highly influence morphology. We run a low mass model (Figure 10), utilising the reference model structural parameters, but fine-tuning SF criteria to account for mass-resolution dependence. Accordingly, we find the peak \asym{I}{} (and the timescale over which \asym{I}{} exceeds the canonical mean of 0.1) is enhanced compared with the reference case. The \asym{I}{} arising in an isolated low-mass model is also exceeds that in more massive models (and greater than previously established limits on secular lopsidedness; B05). In this case therefore, the instantaneous \asym{I}{} is not limited by the corresponding T$_{\rm P}$ (as in the reference model, and the models of B05), with lopsidedness persisting after the driving mechanism has been effectively removed. Also of note is a clear oscillation in the time-varying response of the low-mass model; {\it sloshing} of this nature has been found in previous works (Taga \& Iye 1998), and Athanassoula et al. (2013) attribute oscillations in their models of bar growth to simultaneous non-axisymmetries arising in a triaxial halo.

\subsection{Dependence on SF criteria and feedback}

In models covering a wide parameter range ($e_{\rm SF}$ from 1 to 10 percent, \rhoth{} from 1 to 100 \ncm{}, \fkin{} from 1 to 20 percent and \tsn{} from \ex{4}{} to \ex{5}{} yrs), we find no clear trends with lopsidedness, when quantified in terms of \asym{I}{} and r$_{\rm SF}$. The parameter changes are more often clearly imprinted on other aspects of the spatial distribution: for example, larger values of \rhoth{} promote the formation of a more flocculent gas distribution in the vicinity of the bar ends. A similar arrangement of embryonic arms was noted as common among barred Magellanic systems (NGC 4027, 4618) by de Vacouleurs \& Freeman (1972).

We interpret this result in terms of two competing mechanisms. First, restraining star formation or increasing feedback potency by means of a low $e_{\rm SF}$, high \rhoth{}, low \fkin{} or high \tsn{} mitigates the central mass build-up and bar growth, as quantified by the exponential scalelength and peak m$=$2 amplitude. This is in agreement with previous numerical simulations that reproduce late-type structure (i.e. Brook et al. 2004; Agertz et al. 2011). We find this spiral type exhibits larger peak \asym{I}, bar warp and r$_{\rm SF}$, because the tidal induced lopsided is opposed by less disc self-gravity (Jog 1999), largely as a consequence of a weaker stellar bar.

On the other hand, unrestrained SF and weak feedback tends to promote massive stellar and gas clumps respectively, which also enhance lopsidedness. Figure 10 shows for example that a short timescale for radiative transfer of SNe (\tsn$=$\ex{4}{} yrs) significantly enhances the peak \asym{I}. In these cases, massive clumps scattered through the disc strongly influence the stellar morphology and thus \asym{I}. The clumps heat the disc/halo, weakening bar growth (Noguchi 1996; Athanassoula 2003). This is qualitatively consistent with observations of bluer galaxies, typically late-type spirals, which generally host weaker, smaller bars (i.e. Athanassoula 1992; Hoyle et al. 2011). While these clumps may be argued as unphysical, the gas content and often high SFR of Magellanic-type galaxies draws comparison with galaxies of intermediate redshift, which are also generally quite clumpy (van den Bergh 2002).

\subsection{Dependence on halo and gas fraction}

\begin{figure}
\includegraphics[width=1.\columnwidth]{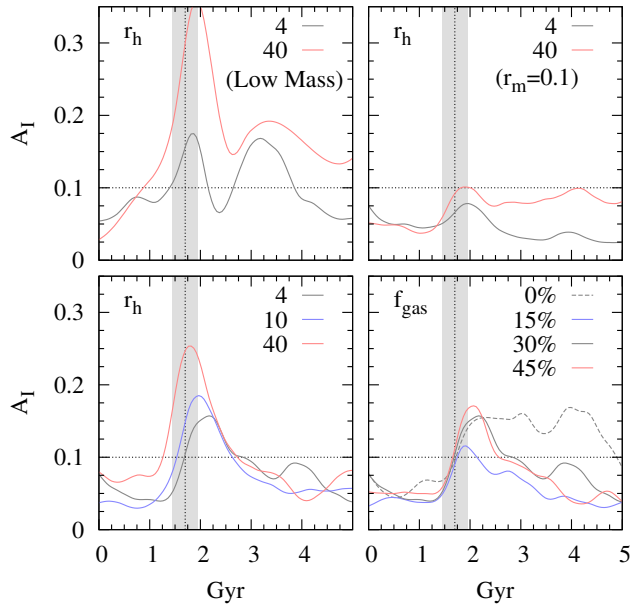}
\caption{Same as Figure 8 for (Top left) Low mass (M=\mex{1.7}{9}{} \md{}) primary with varying halo-to-disc mass ratio (\rh{}); (Top right) One-tenth mass ratio (r$_{\rm m}$$=$0.1) with varying \rh{}; (Bottom left) Reference retrograde model with varying \rh{} (reference \rh{}$=$4); (Right) varying gas mass fraction \fgas{} (reference \fgas{}$=$30 percent)} 
\end{figure}

The LMC's lopsidedness has been interpreted as evidence of a significant dark matter potential (Cioni et al. 2000); external tides can distort the halo, to which the disc responds more strongly than to the perturber alone (Weinberg 1995; Jog 1997). Further evidence for a massive halo in later-type spirals lies in their typically low m$=$2 pattern speeds (Rautiainen, Salo \& Laurikainen 2005). As broached in Section 2.2, estimates for the LMC's dynamical mass are surpassed by theoretical predictions, and we show in Figure 11 that larger r$_{\rm h}$ substantially increase the peak \asym{I}{} (but not the timescale for strong \asym{I}, which remains on the order of a Gyr). For r$_{\rm h}$$=$40, the bar offset for the retrograde case matches the observed 0.5 kpc (van der Marel 2001), as opposed to 0.3 kpc for the reference model with r$_{\rm h}$$=$4.. Moreover, the corresponding r$_{\rm SF}$ and bar warp correlate positively with r$_{\rm rh}$. 

To facilitate fair comparison with our reference model, the halo scaling parameter for these models was fine-tuned to match the inner rotation curve (Figure 3) and bar growth given substantial self-gravity. The bar is accordingly smaller/weaker for larger r$_{\rm h}$ (with no ILR for r$_{\rm h}$$=$40) and the disc scalelength marginally larger. Nonetheless, the results agree with previous simulations; Gao \& White (2007) find larger asymmetry in more massive halos following post-analysis of the Millennium cosmological simulation, and Nordermeer et al. (2001) correlate long-lived lopsidedness with a more halo-dominated inner disc. A similar result is found in two additional cases (Top row, Figure 11): first, in models with one-tenth the mass of the reference case (M$=$\mex{1.7}{9}{} \md{}) \asym{I}{} is increased while maintaining the oscillations noted in Section 3.3. Secondly, a one-tenth mass ratio r$_{\rm m}$ (compared to a reference r$_{\rm m}$$=$0.3) induces a \asym{I}{} barely distinguishable from purely secular origins for the reference \rh{}, but shows long-lived lopsidedness when \rh{}$=$40.

Spirals and dwarves show gas mass fractions ranging from 5 to 80 percent (McGaugh \& de Blok 1997; Geha et al. 2006). We find for non-zero gas mass fractions (\fgas{}) less than the 30 percent used for the reference model,\asym{I}{} increases with \fgas{} (Figure 11). This is consistent with the kinematically colder gas components in the outer disc (as traced by \hi{}) being typically more distorted than the stellar (Richter \& Sancisi 1994). Moreover, a larger gas reservoir can provide a more efficient momentum sink for the bar (Combes \& Elmegreen 1993). This slows its growth and weakens infall that would otherwise lead to a less halo-dominated inner disc, which has been shown to oppose lopsidedness (Jog 1997; Noordermeer et al. 2001). We confirm that the bar is smaller/weaker for larger \fgas{} as found by Athanassoula et al. (2013), and the stellar-to-halo density ratio within the disc scalelength is greatest for lower \fgas{}, although the variation for the latter amounts to only several percent. 

For a collisionless (zero \fgas{}) model, the disc evolution is qualitatively different because the bar strength grows so rapidly without a dissipative component. The inner disc is thus more halo-dominated than our reference model, in accord with simulations of a bar's influence on the halo in collisionless models (Dubinski et al. 2009), which can explain the far longer-lived lopsidedness (Levine \& Sparke 1998). Our results thus suggest multiple mechanisms influencing lopsidedness as an direct or indirect function of \fgas{}. We also note that lopsidedness stemming from a given \fgas{} is sensitive to the corresponding choice of SF/feedback parameters (which are fixed here for simplicity) with regard to the formation of clumps and the changing disc self-gravity (Section 3.4). 

% figure 9 dashed
% figure 5 change scale

\section{Discussion and Conclusions}

In this paper, we investigate how barred late-type spirals are transformed to Magellanic-type galaxies due to tidal interactions. In doing so, we seek to further illustrate the role of tidal interactions in the observed ubiquity of lopsidedness among spirals, with comparison to other proposed mechanisms such as gas accretion. This paper elaborates on previous studies by calibrating our models against the exemplary LMC, and then performing a parameter study to establish the range of lifetimes for lopsidedness, as well as the occurrence of one-sided star formation (OSSF) and disc warps.

A key aspect to our work is the labelling of the distorted galaxy as isolated within 1 to 1.5 Gyr, facilitated by a high-speed tidal encounter. This is comparable to the lifetime for tidal-induced lopsidedness established from previous studies (B05), with the implication that interactions cannot account for the majority of lopsided galaxies. We confirm this lifetime in reference models of the LMC. For a companion galaxy (analogous to the SMC) lying on a retrograde orbit, the morphology and SF history of our reference model matches that of the LMC for up to a Gyr following pericentre. The off-set of the stellar bar from the disc is due largely to distortions of the outer disc, and broadly correlates with stellar lopsidedness (\asym{I}). 

The morphology is qualitatively different between a retrograde and prograde encounters; in particular, out-of-plane distortion caused in the latter is limited to the outer disc. We find systematically more OSSF occurs in the retrograde case, persisting for as long as the bar is also more warped (misaligned to the disc) than in isolated/prograde cases. This is consistent with the recent LMC-SMC being retrograde (Diaz \& Bekki 2012), leading to a warped bar, as implied by observations (Subramaniam 2003) and reproduced in other LMC models (Besla et al. 2012), and with observed OSSF for at least the previous 20 Myr (Harris \& Zaritksy 2009). 

Gas accretion has been recently proposed to account for most lopsidedness among particularly field/isolated galaxies (B05; Mapelli et al. 2008), but it has not yet been established from simulations if such a mechanism can yield OSSF. Magellanic-type galaxies, for which OSSF is characteristic, generally appear isolated (Wilcots \& Prescott 2004), so if tidal interactions are responsible, half would exhibit OSSF, under the reasonable assumption that ret./prograde encounters occur with equal frequency. Indeed, Elmegreen \& Elmegreen (1980) find 50 percent of their Magellanic-type sample exhibit OSSF as indicated by their largest H$_{\rm II}$ regions.

The short-lived \asym{I}{} remains a problem however for the tidal scenario among spirals in general. Moreover, close encounters (including minor mergers) at the mass ratio r$_{\rm m}$$=$0.1 are an order of magnitude more likely than the 0.3 considered for our reference model (Hopkins et al. 2008). The cited work, though considering Milky Way-type massive discs, argues on the premise that the observed low disc thicknesses are a product of a cosmology that favours disc-heating from rare low mass encounters. We find the corresponding \asym{I} for r$_{\rm m}$$=$0.1 to be little more than that induced in secular conditions, and less than predicted from the relationship with T$_{\rm P}$ (i.e. Fig. 11 of B05). This problem could be alleviated by invoking role poorly-detectable dark halos or gas clouds (Bekki 2008; Wilcots et al. 1996), in which case the interaction could be slower (lower $e$) with a correspondingly greater tidal influence. We also cannot rule out the possibility that many companions generally lie at the technological threshold of visibility, as in the recent discovery of the one-tenth mass companion to LMC-analogue NGC 4449 (Martinez-Delgado et al. 2012). 

A less speculative proposal however is that of gas accretion which, in numerical studies of more limited range, is found to yield generally larger peak \asym{I}{} over longer timescales than their respective tidal models (B05; Mapelli et al. 2008), and thus satisfy the observed prevalence of strong \asym{I}. Our broader parameter study finds that the tidal scenario can feasibly match these observations, and address several of the points made in favour of accretion by B05. 

First, B05 find later-type spirals (the precursors to Magellanic-types; Elmegreen et al. 1990) are more lopsided in their OSUBGS survey and suggest accretion can both promote lopsidedness and later-type structure (Bournaud et al. 2005a). Recent studies cast doubt on whether later-types are in fact more lopsided (van Eymeren et al. 2011), although the issue is complicated by environment. The finding that spirals in groups, typically earlier-type, are more lopsided is no surprise given the more frequent harassment. The requirement for high speed encounters is accordingly less however, in which case the tidal forces involved are more effective in converting to earlier types (Mihos \& Hernquist 1994). We find in our high speed interactions no such conversion, such that tidal encounters can both promote \asym{I} and maintain the later-type and lower surface brightness structure more commonly found in fields/isolation. Moreover, later-type spirals typically have more gas (Roberts \& Haynes 1994), and high gas fractions have been shown to mitigate disc conversions during interactions (Hopkins et al. 2009; Martig et al. 2010), as well as promote \asym{I} as shown here.

Second, B05 find a correlation between \asym{I}{} and \asym{II}{} (where using azimuthal Fourier decomposition up to 3.5 scalelength radii) among spiral galaxies, and argue that gas accretion alone can satisfy this correspondence. More recent studies find however an anti-correlation among isolated galaxies (Espada et al. 2011). B05's argument is also motivated by a model of bar renewal via gas accretion (Bournard \& Combes 2002), but the shortening of the bar with time conflicts with observations (Erwin 2005) and simulations of bar growth (e.g. Athanassoula et al. 2013).

In our reference models, \asym{I}{} is most enhanced for the prograde encounter, wherein the bar/spiral is also enhanced, albeit briefly. Our parameter study reveals however that in general, those model prescriptions which promote weaker bars/spirals, facilitate the strongest lopsidedness. Weaker bars are known to correlate with later-type discs (e.g. Abraham \& Merrifield 2000; Erwin 2005) due to the weakened infall of material. The velocity fields of Magellanic-type NGC 925, suggest weak gas funnelling by the bar (Pisano, Wilcots \& Elmegreen 2000). Properties of the LMC are also consistent with a weak bar, including the lack of a gaseous counterpart to the LMC's stellar bar (Besla et al. 2012), and the suspected youth of the bar (Smecker-Hane et al. 2002) which has been possibly stunted in its growth by the persistent tidal heating of the halo. 

Thus Magellanic-types arising from late-type spirals should possess more lopsidedness due to less self-gravity in the inner disc (Jog 1999; Noordermeer et al. 2001). As aforementioned, OSSF is most prominent for more warped bars which, due to their misalignment, constitute weaker dynamical influence with respect to the gas disc. Weaker bars arise in our parameter study for those models which most promote feedback and restrict SF, consistent with recent numerical studies which obtain late-type spirals with realistic halo profiles (Brook et al. 2004; Governato et al. 2010; Agertz et al. 2011). The models of Guedes et al. (2011) suggest adopting a very high resolution ($\sim$10$^{\rm 7}$ particles) is sufficient to resolve late-type structure such as that of the Galaxy, but they use a high \rhoth{} in accord with the mass resolution. It is non-trivial to compare these SF/feedback criteria given their largely phenomenological role, but they appear preferable to our secondary finding that permissive SF or weak feedback can promote weak bars/lopsidedness via the formation of clumps. In future studies, this degeneracy can be avoided by mutual fine-tuning of SF and feedback criteria, to more appropriately model highly localised SF (i.e. Schroyen et al. 2013).

By contrast, we find a clear positive correlation between lopsidedness and the halo-to-disc mass ratio (\rh{}). For models at the LMC stellar mass, there appear to be two mechanisms at play. First, minor distortions in the outer halo can exacerbate luminous asymmetry at the centre (Vesperini \& Weinberg 2000). This is particularly relevant for the common 10:1 mass ratio interaction where the LMC model exhibits lopsidedness only if adopting the \rh{} predicted by halo occupation models (Moster et al. 2010). Second, the more massive halo weakens bar growth (Athanassoula et al. 2013) and, depending on the evolutionary stage of the bar, can facilitate the weak bars we find associated with lopsidedness. The LMC may represent a special case however, where its weak bar may be a consequence of sustained halo heating by successive tidal interactions, rather than amplification through a massive halo which has instead been stripped by the MW (Mastropietro et al. 2005). 

On the other hand, the correspondingly short lifetime of lopsidedness suggests that an increased \rh{} merely amplifies the tidal force rather than sustaining it. In earlier studies, a live halo is speculated to either sustain lopsidedness (as in the case of bars via the removal of angular momentum; Athanassoula 2002) or damp it (Ideta 2002). Levine \& Sparke (1998) find lopsidedness is long-lived if the disc is displaced from the DM potential and spins retrograde to its orbit of the DM centre, but we find little evidence of this occurring in our models, with lopsidedness stemming mainly from distortion in the halo-dominated outer disc (Rix \& Zaritsky 1997). Moreover, the fast tidal encounter prescribed in this work would lead to a fast pattern speed for the global m$=$1 mode, which has been shown analytically to expedite its rapid decline (Ideta 2002). 

We verified in Section 2.5 that the halo profile of our models was acceptably converged to those of higher resolution, but this issue, governed by momentum transport, is notoriously resolution dependent (Valenzuela \& Klypin 2003; Weinberg 1998; Theis 1997), and timely given a parallel debate regarding bar lifetimes (Athanassoula et al. 2002; 2013). Furthermore, the self-consistent treatment of extant halo triaxiality has not been studied in these contexts to date. Athanassoula et al. (2013) find a triaxial halo stabilizes against bar growth, in which case lopsidedness would be enhanced according to the conclusions of this work, although distortions incurred in the interaction itself remains to be accounted for. 

We find lower mass galaxies exhibit stronger lopsidedness, and are particularly responsive to amplification by more massive halos. \asym{I} often conveys oscillations due to the complex coupling of non-asymmetries in a triaxial halo (Athanassoula et al. 2013). One implication is that tidally disturbed discs could be mis-identified as not lopsided galaxy if observed during a {\it trough} in this periodicity. More significantly, the timescale of lopsidedness can persist to several Gyr, comparable to the models of gas accretion (B05). In support of accretion, B05 argue that late-types are typically less massive and thus will be more lopsided for a given mass of accreted gas. In analytical models of star forming galaxies, however, Bouche et al. (2010) match a universal mass-dependence on the condition that gas accretion for fuelling star formation is greatly truncated below a halo mass floor at $\sim$\ex{11}{} \md{} due to photoionization effects. Thus we postulate that tidal interactions are a significant progenitor of lopsidedness towards lower mass, but encourage future studies to explore this mass dependence.

\section*{Acknowledgements}

CY is supported by the Australian Postgraduate Award Scholarship. We gratefully acknowledge the suggestions of the referee that improved this manuscript.

\bsp
\label{lastpage}
\end{document}